\documentclass{ws-procs9x6}

\begin{document}

\title{VECTOR AND AXIAL-VECTOR CURRENT CORRELATORS WITHIN THE INSTANTON MODEL OF QCD VACUUM.}

\author{A.E. DOROKHOV}

\address{Bogoliubov Laboratory of Theoretical Physics,\\
Joint Institute for Nuclear Research, \\
141980, Dubna, Moscow Region, Russia.\\
E-mail: dorokhov@thsun1.jinr.ru}

\maketitle

\abstracts{
The pion electric polarizability, $\alpha^E _{\pi ^{\pm }}$, the leading order hadronic contribution
to the muon anomalous magnetic moment, $a_{\mu}^{\mathrm{hvp}\left(  1\right)}  $,
and the ratio of the difference to the sum of vector and axial-vector
correlators, $(V-A)/(V+A)$,
are found within the instanton model of QCD vacuum. The results are
compared with phenomenological estimates of these quantities following from the ALEPH
and OPAL data on vector and axial-vector spectral densities.}

In the chiral limit, where the masses of $u$, $d$, $s$ light quarks are set to
zero, the vector ($V$) and axial-vector ($A$) current-current correlation
functions in the momentum space (with $-q^{2}\equiv Q^{2}\geq0$) are defined
as
\begin{eqnarray}
\Pi_{\mu\nu}^{J,ab}(q) &  =&i\int d^{4}x~e^{iqx}\Pi_{\mu\nu}^{J,ab}%
(x)=\,\left(  q_{\mu}q_{\nu}-g_{\mu\nu}q^{2}\right)  \Pi_{J}(Q^{2})\delta
^{ab},\label{PA}\\
\qquad\Pi_{\mu\nu}^{J,ab}(x) &  =&\langle0\left\vert T\left\{  J_{\mu}%
^{a}(x)J_{\nu}^{b}(0)^{\dagger}\right\}  \right\vert 0\rangle,\nonumber
\end{eqnarray}
where the QCD $V$ and $A$ currents are
$
J_{\mu}^{a}=\overline{q}\gamma_{\mu}\frac{\lambda^{a}}{\sqrt{2}}q,\
J_{\mu}^{5a}=\overline{q}\gamma_{\mu}\gamma_{5}\frac{\lambda^{a}}{\sqrt{2}%
}q,\label{JAV}%
$
and $\lambda^{a}$ are Gell-Mann matrices $\left(  \mathrm{tr}\lambda
^{a}\lambda^{b}=2\delta^{ab}\right)$ in flavor space. The momentum-space two-point
correlation functions obey (suitably subtracted) dispersion relations,
\begin{equation}
\Pi_{J}(Q^{2})=\int_{0}^{\infty}\frac{ds}{s+Q^{2}}\frac{1}{\pi}\mathrm{Im}%
\Pi_{J}(s),\label{Peuclid}%
\end{equation}
where the imaginary parts of the correlators determine the spectral functions
$
\rho_{J}(s)=4\pi{\mathrm{Im}}\Pi_{J}(s+i0)\nonumber
$ measured by ALEPH \cite{ALEPH2} and OPAL \cite{OPAL}.

In the instanton liquid model (ILM) gauged by interaction with external vector
and axial-vector fields \cite{ADoLT00} the correlators in the chiral limit have transverse character
\cite{Dorokhov:2003kf,Dorokhov:2004pw}
\begin{equation}
\Pi_{\mu\nu}^{J}\left(  Q^{2}\right)  =\left(  g_{\mu\nu}-\frac{q^{\mu}q^{\nu
}}{q^{2}}\right)  \Pi_{J}^{\mathrm{ILM}}\left(  Q^{2}\right)
,\label{PVmn}%
\end{equation}
and the dominant contribution to the correlators is given by the
dynamical quark loop which was found in \cite{Dorokhov:2003kf,Dorokhov:2004ze} with the
result for the vector current
\begin{eqnarray}
&&\Pi_{V}^{Q\mathrm{Loop}}\left(  Q^{2}\right)   =\frac{4N_{c}}{Q^{2}}%
\int\frac{d^{4}k}{\left(  2\pi\right)  ^{4}}\frac{1}{D_{+}D_{-}}\left\{
M_{+}M_{-}+\left[  k_{+}k_{-}-\frac{2}{3}k_{\perp}^{2}\right]  _{ren}\right.
\label{Ploop}\\
&  +&\left.  \frac{4}{3}k_{\perp}^{2}\left[  \left(  M^{\left(  1\right)
}\left(  k_{+},k_{-}\right)  \right)  ^{2}\left(  k_{+}k_{-}-M_{+}%
M_{-}\right)  -\left(  M^{2}\left(  k_{+},k_{-}\right)  \right)  ^{\left(
1\right)  }\right]  \right\}  +\nonumber\\
&  +&\frac{8N_{c}}{Q^{2}}\int\frac{d^{4}k}{\left(  2\pi\right)  ^{4}}%
\frac{M\left(  k\right)  }{D\left(  k\right)  }\left[  M^{\prime}\left(
k\right)  -\frac{4}{3}k_{\perp}^{2}M^{\left(  2\right)  }\left(
k,k+Q,k\right)  \right]  ,\nonumber
\end{eqnarray}
where the notations
$
k_{\pm}=k\pm Q/2, k_{\perp}^{2}=k_{+}k_{-}-\frac{\left(  k_{+}q\right)
\left(  k_{-}q\right)  }{q^{2}}, D\left(  k\right)  =k^{2}+M^{2}(k),
$
and $
M_{\pm}=M(k_{\pm}),\  D_{\pm}=D(k_{\pm})
$
are used. We also introduce the finite-difference derivatives defined for an
arbitrary function $F\left(  k\right)  $ as
\begin{equation}
F^{(1)}(k,k^{\prime})=\frac{F(k^{\prime})-F(k)}{k^{\prime2}-k^{2}},\qquad
F^{(2)}\left(  k,k^{\prime},k^{\prime\prime}\right)  =\frac{F^{(1)}%
(k,k^{\prime\prime})-F^{(1)}(k,k^{\prime})}{k^{\prime\prime2}-k^{\prime2}}.
\label{FDD}%
\end{equation}

The difference of the $V$ and $A$ correlators is free from any perturbative corrections
for massless quarks and very sensitive to the spontaneous breaking of chiral symmetry.
The model calculations of the chirality flip $V-A$ combination provides
\begin{eqnarray}
&&\Pi_{V-A}^{Q\mathrm{Loop}}\left(  Q^{2}\right)   =-\frac{4N_{c}}{Q^{2}}%
\int\frac{d^{4}k}{\left(  2\pi\right)  ^{4}}\frac{1}{D_{+}D_{-}}\left\{
M_{+}M_{-}+\frac{4}{3}k_{\perp}^{2}\cdot
\right.  \label{VmAmodel}\\
&&\left.
\left[  -\sqrt{M_{+}M_{-}}M^{\left(
1\right)  }\left(  k_{+},k_{-}\right)
 +\left(  \sqrt{M}^{\left(  1\right)  }\left(  k_{+}%
,k_{-}\right)  \right)  ^{2}\left(  \sqrt{M_{+}}k_{+}+\sqrt{M_{-}}%
k_{-}\right)  ^{2}\right]  \right\}  .\nonumber
\end{eqnarray}
Vice versa, $V$ and $A$ correlators are separately dominated by perturbative
massless quark loop diagram in the high momenta region. In the model calculations this
dominance is reproduced because in the chiral limit the dynamical quark mass generated
in the instanton vacuum, $M(k)$, vanishes at large virtualities $k^2$.

\begin{figure}[h]
\hspace*{-0cm}
\vspace*{0.0cm} \epsfxsize=7cm \epsfysize=4.5cm \centerline{\epsfbox
{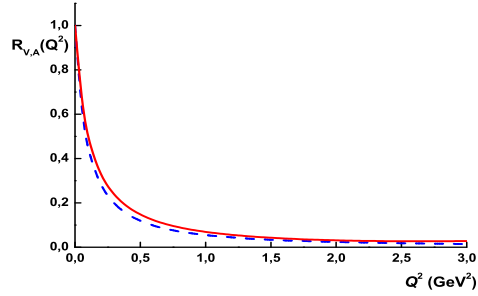}}
\caption[dummy0]{Normalized $V-A$ correlation function Eq. (\ref{Rcorr})
obtained from the model (solid) and
reconstructed from the ALEPH experimental spectral functions
with next-to-leading accuracy in $\alpha_s$ (dashed).}%
\label{Fv-a}
\end{figure}

With help of the Das-Mathur-Okubo (DMO) sum rule it is possible to estimate
the electric polarizability of the charged pions by using the Gerasimov relation
\begin{equation}
\alpha^E _{\pi ^{\pm }}=\frac{\alpha }{m_{\pi }}\left[ \frac{\left\langle
r_{\pi }^{2}\right\rangle }{3}-\frac{I_{DMO}}{f_{\pi }^{2}}\right] ,
\label{PiPolariz}
\end{equation}%
where $I_{DMO}$ is the integral corresponding to the DMO sum rule
\begin{equation}
I_{DMO} =\frac{1}{4\pi ^{2}}\int_{0}^{\infty}\frac{ds}{s}%
\left[ \rho_V\left( s\right) -\rho_A\left( s\right) \right] =\left. \frac{\partial }{%
\partial Q^{2}}\left[ Q^{2}\Pi _{V-A}\left( Q^{2}\right) \right]
\right\vert _{Q^{2}\rightarrow 0}.  \label{Idmo}
\end{equation}%
From (\ref{PiPolariz}) with values of charged pion radius and $I_{DMO}$ obtained from
the model (see for further details \cite{Dorokhov:2004ze,DRV03}) one finds the value
\begin{equation}
\left[ \alpha^E _{\pi ^{\pm }}\right] _{\mathrm{model}}=2.9\cdot 10^{-4}\mathrm{fm}%
^{3},  \label{AlMod}
\end{equation}%
which is close to experimental numbers
$
\left[  \alpha_{\pi^{\pm}}^{E}\right]  _{\mathrm{exp}}^{\mathrm{OPAL}}=
2.71(88)  \cdot10^{-4}\mathrm{~fm}^{3}$
\cite{OPAL} and \cite{PIBETA}
$\left[  \alpha_{\pi^{\pm}}^{E}\right]  _{\mathrm{exp}}^{\mathrm{PIBETA}%
}=\left\{
\begin{array}
[c]{l}%
2.68(9)\cdot10^{-4}\mathrm{~fm}^{3}\qquad\mathrm{{full\ data\ set},}\\
2.90(9)\cdot10^{-4}\mathrm{~fm}^{3}\qquad
\mathrm{kinematically\ restricted\ data\ set}.
\end{array}
\right.$  New precise results on pion and kaon polarizabilities are expected from COMPASS
\cite{Bressan}.

The leading order hadronic
vacuum contribution to the lepton anomalous magnetic moment is given by
\begin{equation}
a_{\mu}^{\mathrm{hvp}\left(  1\right)  }=-\frac{2}{3}\alpha^{2}\int_{0}%
^{1}dx
\left(  1-x\right)\bar \Pi_{V}\left(
\frac{x^{2}}{1-x}m_{l}^{2}\right)  ,\label{aAd}%
\end{equation}
where $\bar \Pi_V(Q^2)=\Pi_V(Q^2)-\Pi_V(0)$, $m_{l}$ is the lepton mass,
and the charge factor $2/3$ is taken into
account. One gets the model estimate
\begin{equation}
[a_{\mu}^{\mathrm{hvp}\left(  1\right)  }]_{\mathrm{model}}=6.2\left(
0.4\right)\cdot10^{-8}\label{ammALEPH}%
\end{equation}
which is in a reasonable agreement with the phenomenological numbers,
found from precise determination of the low energy tail of the total $e^{+}
e^{-}\rightarrow$ hadrons and $\tau$ lepton decays cross-sections\cite{blr04}
\begin{equation}
[a_{\mu}^{\mathrm{hvp}\left(  1\right)  }]_{\mathrm{exp}}=\left\{
\begin{array}
[l]{l}%
6.934(88)\cdot10^{-8},\qquad e^{+}e^{-},\\
7.018(90)\cdot10^{-8},\qquad\tau.
\end{array}
\right.
\end{equation}
As by product we estimate also the anomalous magnetic moment of the $\tau$ lepton as
\cite{Dorokhov:2004ze}
$
[a_{\tau}^{\mathrm{hvp}\left(  1\right)  }]_{\mathrm{model}}=3.1\left(0.2\right)\cdot10^{-6}.
$

The ratio of correlators
\begin{equation}
R_{V,A}(Q^2)=\frac{\bar \Pi_{V-A}\left(  Q^{2}\right) }{\bar \Pi_{V-A}\left(  Q^{2}\right) -
2\bar \Pi_{V}\left(Q^{2}\right) }
\label{Rcorr}\end{equation}
characterizes the chirality transfer in dependence of passing virtuality. In (\ref{Rcorr})
the vector correlator $\bar \Pi_{V}\left(Q^{2}\right)$ is defined above and the
axial-vector correlator is defined with kinematical pole removed:
$\bar \Pi_{V-A}(Q^2) =  \Pi_{V-A}(Q^2)+f^2_\pi/Q^2$. This pole is not visible in experiment
and difficult for detection on the lattice.
From general grounds
one expects that this ratio is unit at zero virtuality, and it goes to zero at large virtualities
where perturbative dynamics dominates.
In Fig.~1 we present the ratio of correlators, $R_{V,A}$, reconstructed from ALEPH spectral data.

The region of intermediate momentum transfer provides nontrivial
transition between low energy dynamics described in terms of chiral symmetry structures and
high energy dynamics with relevant operator product expansion language. The problems of standard
approaches are the rapid growth of independent operator structures with less sensitivity of their
experimental determination when moving beyond the
applicability region. Moreover, there is a problem to define an energetic
scale at which the standard expansions begin to work. These problems can be overcome with the aid of
the instanton model of QCD vacuum. New data from ALEPH and OPAL on inclusive hadronic $\tau$ lepton
decays are very helpful in study of current correlators, one of the simplest objects,
at intermediate momentum transfer.

\section*{Acknowledgments}
The author thanks the Organizers of Spin 2004 for hospitality
and financial support. The work is partially supported by RFBR
(Grant nos. 04-02-16445, 03-02-17291, 02-02-16194) and by Heisenberg-Landau program.

\end{document}